# A low-field temperature-dependent EPR signal in terraced $MgO:Mn^{2+}$ nanoparticles: an enhanced Zeeman splitting in the wide-bandgap oxide

Peter V. Pikhitsa[a)], Sukbyung Chae, Seungha Shin, and Mansoo Choi[a)]

*Global Frontier Center for Multiscale Energy Systems, Division of WCU Multiscale Mechanical Design , School of Mechanical and Aerospace Engineering, Seoul National University, Seoul, Korea, 151-742*

$Mn^{2+}$ ion doping is used as an electron paramagnetic resonance (EPR) probe to investigate the influence of low-coordination structural defects such as step edges at the surface of terraced (001) MgO nanoparticles on the electronic properties. Beside the well-known hyperfine sextet of $Mn^{2+}$ ions in the cubic crystal field of MgO, an additional EPR feature with a striking non-monotonous temperature dependent shift of the g-factor is observed in terraced nanoparticles in the temperature range from 4K to room temperature. By linking the difference in the temperature dependence of the $Mn^{2+}$ sextet intensity in cubic and terraced nanoparticles with the possible s-d exchange shift and enhanced Zeeman splitting we conclude that the novel EPR feature originates from the loosely trapped charge-compensating carriers at the abundant structural defects at the surface of terraced nanoparticles due to their exchange interaction with neighboring $Mn^{2+}$ ions.

[a)] Authors to whom correspondence should be addressed. Electronic mail: peter@snu.ac.kr, mchoi@snu.ac.kr

Iconic wide-gap oxide MgO has been used for many applications connected with the spin states of electrons, such as spintronics and superconductivity, mostly for buffer tunneling layers in nanoscale. On the other hand, magnetic $Mn^{2+}$ ions, that play a decisive role in spintronics of diluted magnetic semiconductors (DMS), can be easily doped into MgO and its nanostructures. Pure MgO forms an ionic crystal with cubic NaCl structure consisting of $Mg^{2+}$ and $O^{2-}$. On contrast, the chemical bond in a free MgO molecule is much more covalent than in the bulk of the crystal and the effective charges on the ions are $Mg^{1+}$ and $O^{1-}$. Thus, in the bulk of MgO crystal the electron affinity is practically eliminated by the Madelung potential. Meanwhile, (001) MgO surface can have structural defects such as step edges, kinks, corners *etc.* that can possess various electronic properties interesting for applications such as sensors and catalysts. Having the band-gap of 7.7 eV, completely filled valence bands, and empty conduction bands, MgO would transfer electrons/holes only along the low-coordination structural surface defects – mostly step edges. It could happen because the Madelung potential decreases rapidly as the Mg and O coordination decreases, and for the Mg ion at the edge, corner, and at other low coordinated (LC) sites the electron affinity becomes comparable with the electron affinity for a free ion. The calculated relaxed electron affinities of 3-coordinated Mg sites of 1-2 eV demonstrate that they can serve as electron traps. [1, 2] A trapped electron is almost entirely localized at the terminating Mg ion, which thus becomes similar to $Mg^{1+}$. The same is true for trapped holes and for $O^{1-}$ ions. One would expect that foreign ions like $Mn^{2+}$ if close enough to the LC sites would also need charge compensation like $Mn^{2+} + e^-$ to produce an electron-$Mn^{2+}$ complex.

For MgO nanoparticles the surface structural defects depend on the method of synthesis[3] that may also produce bulk defects. Recently we reported that terraced MgO nanoparticles are distinguished by peculiar bulk electronic defect states leading to luminescence with a double-band light emission at 260 nm and 490 nm.[3] A further attempt to investigate electronic states in terraced MgO nanoparticles with the Electron Paramagnetic Resonance (EPR) technique at the X-band by comparing them with cubic MgO nanoparticles led us to unexpected results. In fact, as far as the Inductively Coupled Plasma analysis indicated that all the nanoparticles always contain unintentionally doped $Mn^{2+}$ impurities at a several ppm level,[3] a pronounced sextet signal from those impurities was only expected to be identical for cubic and terraced nanoparticles.

Using $Mn^{2+}$ as an EPR probe has been well-known.[4-6] On the other hand, numerous papers report optical detection that demonstrates how a single $Mn^{2+}$ ion influences the DMS exciton through the sp-d exchange field in a carrier-Mn complex in semiconductor single quantum dots.[7-9] The shift in the band edge energy leads to a great $g$-factor enhancement.[10-12] If LC defects trapped a charge carrier (as far as such a charge transfer was predicted for Mg and O surface ions) [1, 2] then an $Mn^{2+}$ ion would also influence the nearby electron through the s-d exchange field.

Indeed, for both cubic and terraced nanoparticles we found the well-known hyperfine sextet of $Mn^{2+}$ ions, usually for (001) MgO, yet for terraced ones we observed an additional broad feature with $g$-factor of 2.23 at room temperature (RT). Quite puzzling, this feature moves downfield considerably when the temperature lowers to 4 K. Such lability makes it difficult to assign this feature to an isolated paramagnetic defect. The $g$ factor has a non-monotonous temperature dependence that may be interpreted as a shift from the electron free value, that may be caused by the exchange field of nearby $Mn^{2+}$ ions. Meanwhile, a low-field $g$=2.77 feature has been known in GaAs: $Mn^{2+}$ and was explained as the signal from the complex $Mn^{2+}$ ($d^5$) plus the weakly coupled delocalized hole.[13] Note that an excited paramagnetic system in a triplet state with the exchange already showed a striking and unexplained non-monotonous $g$ shift (similar to what we observed) proportional to a quantity differing only slightly from the susceptibility.[14]

Here we report the results of our EPR measurements on terraced and cubic MgO nanoparticles and correlate the temperature behavior of the low-field EPR feature with the temperature behavior of the EPR signal from $Mn^{2+}$ ions. By this correlation we conclude that the new EPR feature of terraced MgO nanoparticles may originate from trapped electrons/holes that were predicted to exist along the edges and ledges of the terraces being LC structural defects. [1, 2] The electrons may exchange-interact with nearby $Mn^{2+}$ ions which provide an effective field that acts on the loosely trapped electrons as the EPR shows. Thus the puzzling EPR feature has been qualitatively explained and its existence can be considered as supporting the presence of loosely bound charge carriers along the edges of terraces that may find an application in nanoelectronics, based on the wide-gap oxide.

MgO nanoparticles were produced in an oxy-hydrogen flame and were distinguished by intensive terraced structure (say, of type I, Fig.1(b)) as described earlier in Ref.[3] Type II nanoparticles (Fig.1(a)) were produced by burning Mg in dry air and were distinguished by exclusively cubic shape of nanoparticles (well-known MgO smoke). The powder samples were measured in a standard Bruker EPR spectrometer setup equipped with a helium flow cooling system with temperature stabilization. The measurements were performed in a sweeping mode in X-band of about 9.4 GHz. Both types of nanoparticles contained similar amount of $Mn^{2+}$ impurities of several tens ppm revealed by EPR as the characteristic sextet of hyperfine splitting (HFS) of $Mn^{2+}$ ions in the cubic crystal field (Fig.2) with $g$-factors of (1.88; 1.93; 1.98; 2.03; 2.09; 2.14) . The sextet is known to have been widely used as a marker along with the famous DPPH for calibration of $g$-factors in EPR experiments. However, only for terraced samples of type I there appeared an additional broad EPR signal (temperature-dependent line-widths of ca. 100-200 Oe) at low field with $g = 2.23$ at RT that was not observed for type II samples. Moreover, the additional low-field EPR feature demonstrated a complex behavior with the temperature (Fig. 3). The enhanced $g$-factor extracted from graphs in Fig. 3 is shown in Fig. 4 and is pronouncedly temperature dependent when $g$-factor changed up to $g = 2.7$ at 4 K (Fig.4). In parallel, one can observe a non-Curie temperature dependence of the HFS sextet intensity for both types of the samples (Fig.2), however, the specific temperature dependences were quite different for type I and type II samples as presented in Fig.2 (b) and (d), correspondingly.

A non-Curie behavior seems quite natural for the $Mn^{2+}$ ion sextet because of the interplay between the thermal and radio-frequency energy population of the energy levels.[15] However, the difference between Fig. 1(b) and (c) suggests that some $Mn^{2+}$ ions in type I samples participate in the resonance differently just because they may lie in the vicinity of the LC structural defects that may modify the charge state of the surrounding oxygen ion. Therefore, as we mentioned before, there is a need in charge compensation, so similar to $Mg^{1+}$ ion on the surface, an $Mn^{1+}$ ion could be created. However, if all ions were such, the $Mn^{1+}$ ions (being a $d^6$ non-Kramers ion) would be nearly EPR silent. Thus instead one may rather expect a number of complexes electron -$Mn^{2+}$, such that the electron may be donated to a nearby LC defect and be loosely bound.

Analogous charge compensation by a hole was discussed in Ref.[8] It is intriguing to notice that in GaAs: Mn the bound hole produced a broad low-field EPR feature at $g$=2.77 along with the $Mn^{2+}$ signal.[13] Additionally, a flip-flop in two resonant forms $Fe^{3+}$----$O^{2-}$ ↔ $Fe^{2+}$----$O^{1-}$ (it could be realized in the $Mn^{2+}$---$O^{1-}$ state jumping into the $Mn^{3+}$---$O^{2-}$ state and back) was reported in Ref.[16] as the main reason for EPR feature (the radiation dose dependent) at $g \approx 2.54$ in feldspars.

We suggest a simple semi-quantitative analysis of our experimental findings based on the analogy with DMS. A loosely trapped electron brought to an LC defect is in the field of a nearby $Mn^{2+}$ ion which magnetic moment relaxes relatively slow. The local field induced by the exchange between the electron and the ion influences the carrier Ref.[12] The local field from the magnetic ion makes magnetization $M = \chi(T)H$ leading to a shift in a resonance field of the nearby loosely trapped electron, thus producing a temperature dependent $g$-factor. It is similar to what happens when an $Mn^{2+}$ ion acts on an exciton in a quantum dot with a so-called giant Zeeman effect.[10-12]

We can write down the equation for the effective $g$-factor of the electron that explicitly expresses it through magnetization and the susceptibility in analogy with Ref.[12]:

$$g = g_e + \frac{\alpha M}{g_{Mn}\mu_B{}^2 H} = 2 + \frac{\alpha\chi}{2\mu_B{}^2}\ , \qquad (1)$$

where $g_e \approx 2$ is the loosely trapped electron $g$-factor (this is quite different from the DMS where instead of $g_e$in Eq.(1) there is $g^*$ - the negative $g$-factor of the band electron) and $g_{Mn} \approx 2$ is the $Mn^{2+}$ $g$-factor, $\alpha$ is the exchange integral, positive for ferromagnetic interaction between the electron and $Mn^{2+}$ ion, $\mu_B$ is the Bohr magneton, and $\chi$ is the static magnetic susceptibility.

Instead of $\chi$, in Eq. (1) we deal with the EPR susceptibility $\chi(T)$ which is proportional to the EPR response of the $Mn^{2+}$ ions residing at the LC defects and which can be obtained by subtracting the appropriately rescaled EPR susceptibility $\chi_C(T)$ of cubic samples (calculated in the usual way from the linewidth and the intensity of the signal in arbitrary units) taken from Fig. 2(b), from the EPR susceptibility $\chi_T(T)$ of terraced samples (taken from Fig.2(d)):

$$\chi(T) = \chi_T(T) - A\chi_C(T), \quad (2)$$

where $A > 0$ is a constant. Finally, the $g$-factor is the sum of the free electron value and the shift

$$g(T) = 2 + B[\chi_T(T) - A\chi_C(T)] \ , \quad (3)$$

where $B > 0$ is another constant. Both $A$ and $B$ can be obtained from comparing $g(T)$ with the experimental dependence in Fig. 4. The line with the fitting parameters $A = 0.25$ and $B = 7 \times 10^{-5}$ is rather close to the experimental data. Finally, it also qualitatively reproduces the otherwise puzzling non-monotonous behavior of $g(T)$.

In conclusion, we presented unusual EPR data of a nanostructured material, based on MgO terraced nanoparticles, that demonstrated a possibility of manipulating the electronic properties of the wide-gap oxide by magnetic $Mn^{2+}$ ions similar to the well-established manipulation with DMS quantum dots. An encouraging example of such manipulation is room temperature magnetism found in ZnO:$Mn^{2+}$ Ref. [17]

This work was done by financial support from the Global Frontier Center for Multiscale Energy Systems (2011-0031561) supported by the Korean Ministry of Science and Technology. We thank Korea University and Sunhee Kim from KBSI for the EPR measurements.

## REFERENCES

[1] M. Chiesa, M. C. Paganini, E. Giamello, D. M. Murphy, C. Di Valentin, and G. Pacchioni, Acc. Chem. Res. **39**, 861 (2006).

[2] K. P. McKenna, P. V. Sushko, and A. L. Shluger, J. Am. Chem. Soc. **129**, 8600 (2007).

[3] P. V. Pikhitsa, C. Kim, S. Chae, S. Shin, S. Jung, M. Kitaura, S. Kimura, K. Fukui, and M. Choi, Appl. Phys. Lett. **106**, 183106 (2015).

[4] T. Yamamura, A. Hasegava, Y. Yamada, and M. Miura, Bull. Chem. Soc. Japan, **43**, 3377 (1970).

[5] E. G. Derouane, J. Thelen, and V. Indovina, Bull. Soc. Chim. Belg. **82**, 657 (1973).

[6] T. Story, C. H. W. Swüste, P. J. T. Eggenkamp, H. J. M. Swagten, and W. J. M. de Jonge, Phys. Rev. Lett. **77**, 2802 (1996).

[7] M. Goryca, M. Koperski, T. Smolenski, Ł. Cywinski, P. Wojnar, P. Plochocka, M. Potemski, and P. Kossacki, Phys. Rev. B **92**, 045412 (2015).

[8] A. Kudelski, A. Lemaıtre,1, A. Miard, P. Voisin, T. C. M. Graham, R. J. Warburton, and O. Krebs1, Phys. Rev. Lett. **99,** 247209 (2007).

[9] L. Besombes, Y. Leger, L. Maingault, D. Ferrand, H. Mariette, and J. Cibert, Phys. Rev. Lett. **93**, 207403 (2004).

[10] J. A. Gaj, R. Planel, and G. Fishman, Solid State Com. **29**, 435 (1979).

[11] F. J. Teran, M. Potemski, D. K. Maude, D. Plantier, A. K. Hassan, A. Sachrajda, *et al*, Phys. Rev. Lett. **91**, 077201 (2003).

[12] J. K. Furdyna, J. Appl. Phys. 64, R29 (1988).

[13] J. Schneider, U. Kaufmann, W. Wilkening, M. Baeumler, and F. Kohl, Phys. Rev. Lett. **59**, 240 (1987).

[14] B. R. Cooper, R. C. Fedder, and D. P. Schumacher, Phys. Rev. **163**, 506 (1967).

[15] M. Godlewski, S. Yatsunenko, and V. Yu. Ivanov, Israel J. Chem, **46**, 413 (2006).

[16] M. D. Sastry, Y. C. Nagar, B. Bhushan, K. P. Mishra, V. Balaram and A. K. Singhvi, J. Phys.: Condens. Matter, **20,** 025224 (2008).

[17] N. S. Norberg, K. R. Kittilstved, J. E. Amonette, R. K. Kukkadapu, D. A. Schwartz, and D. R. Gamelin, J. Am. Chem. Soc. **126**, 9387 (2004).

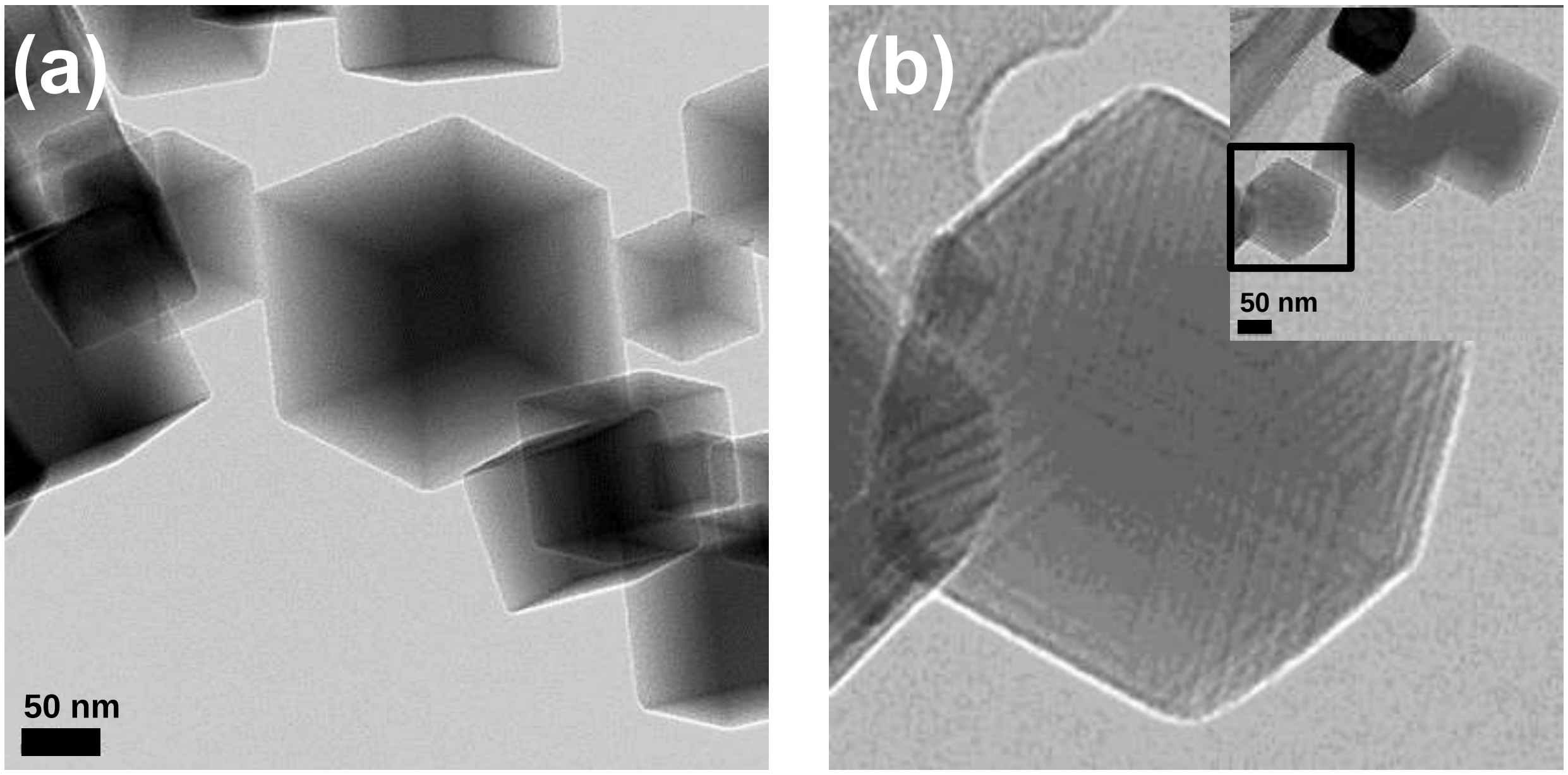

FIG.1. TEM images of type II (a) and type I (b) MgO nanoparticles. In (b) the pronounced LC structural defects are clearly seen.

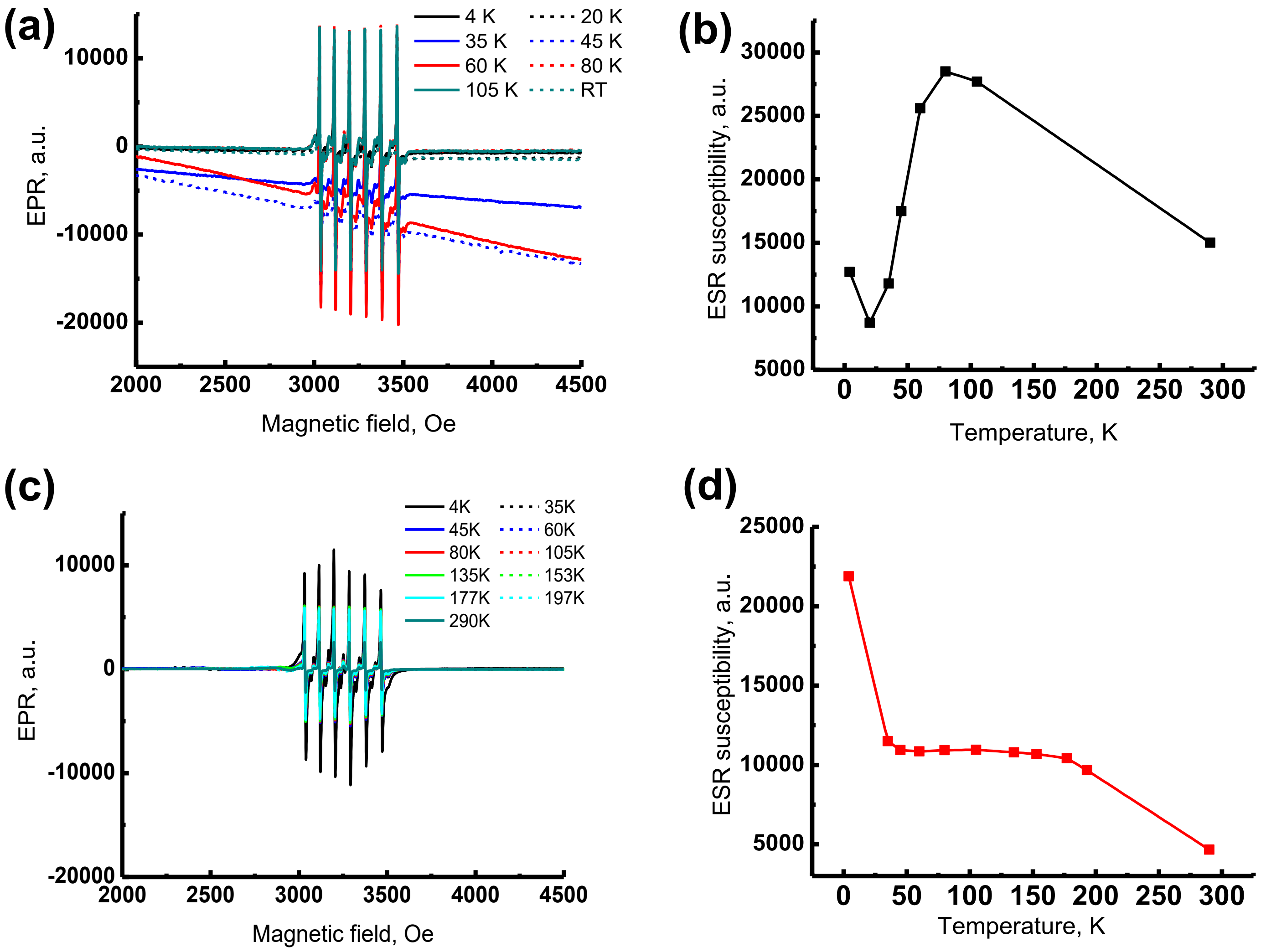

FIG. 2. (Color online). EPR spectra of type II (a) and type I (c) MgO nanoparticles. (b), (d), temperature dependencies of the EPR intensity of type II and type I nanoparticles, respectively.

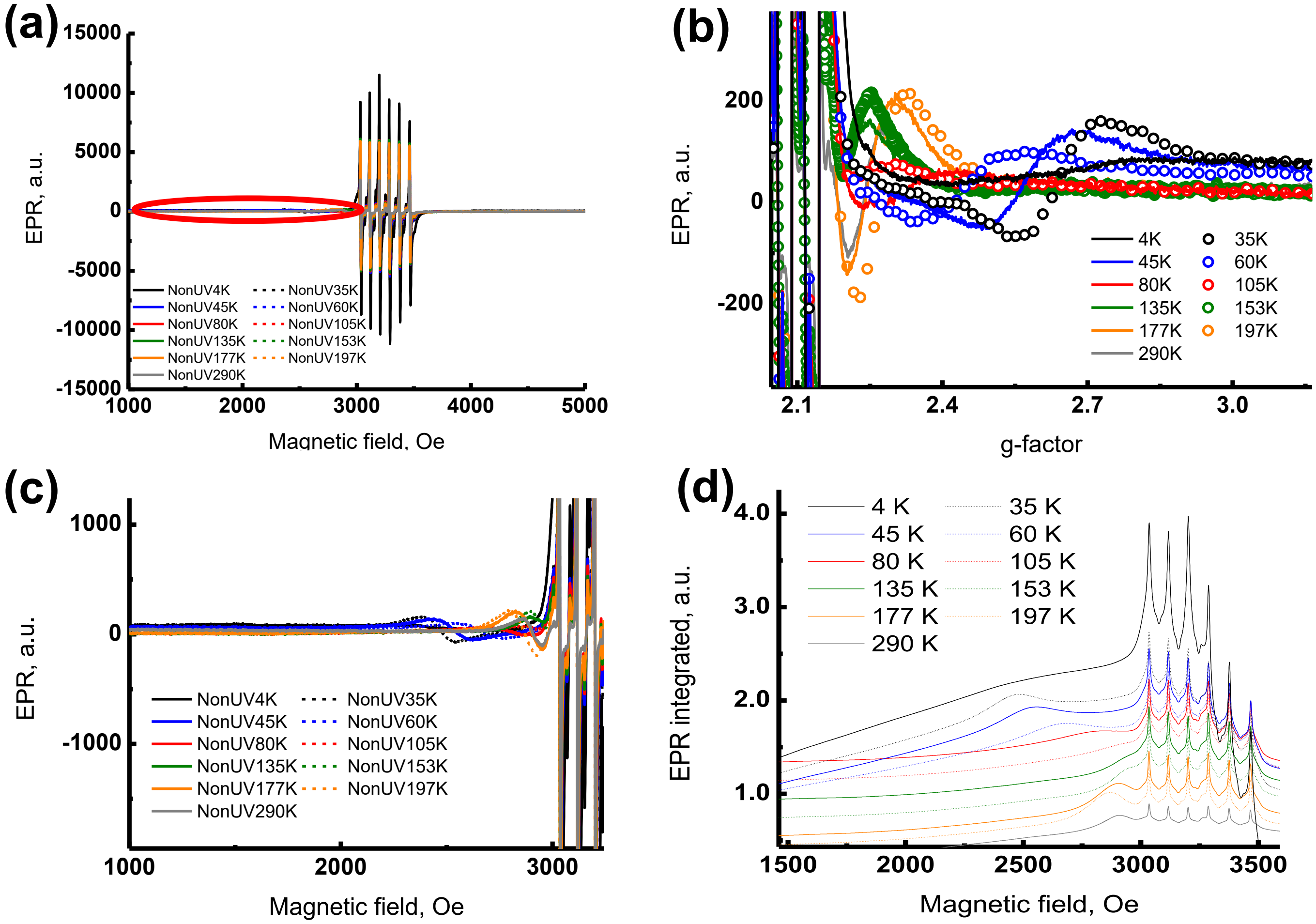

FIG. 3. (Color online). EPR of type I MgO nanoparticles measured at different temperatures. (a) the EPR spectra, (c) a magnified region encircled in (a), (b) re-plotted (c) with the $g$-factor instead of the field, (d) EPR absorption as the integrated signal from (c). The lines are vertically shifted for better comprehension.

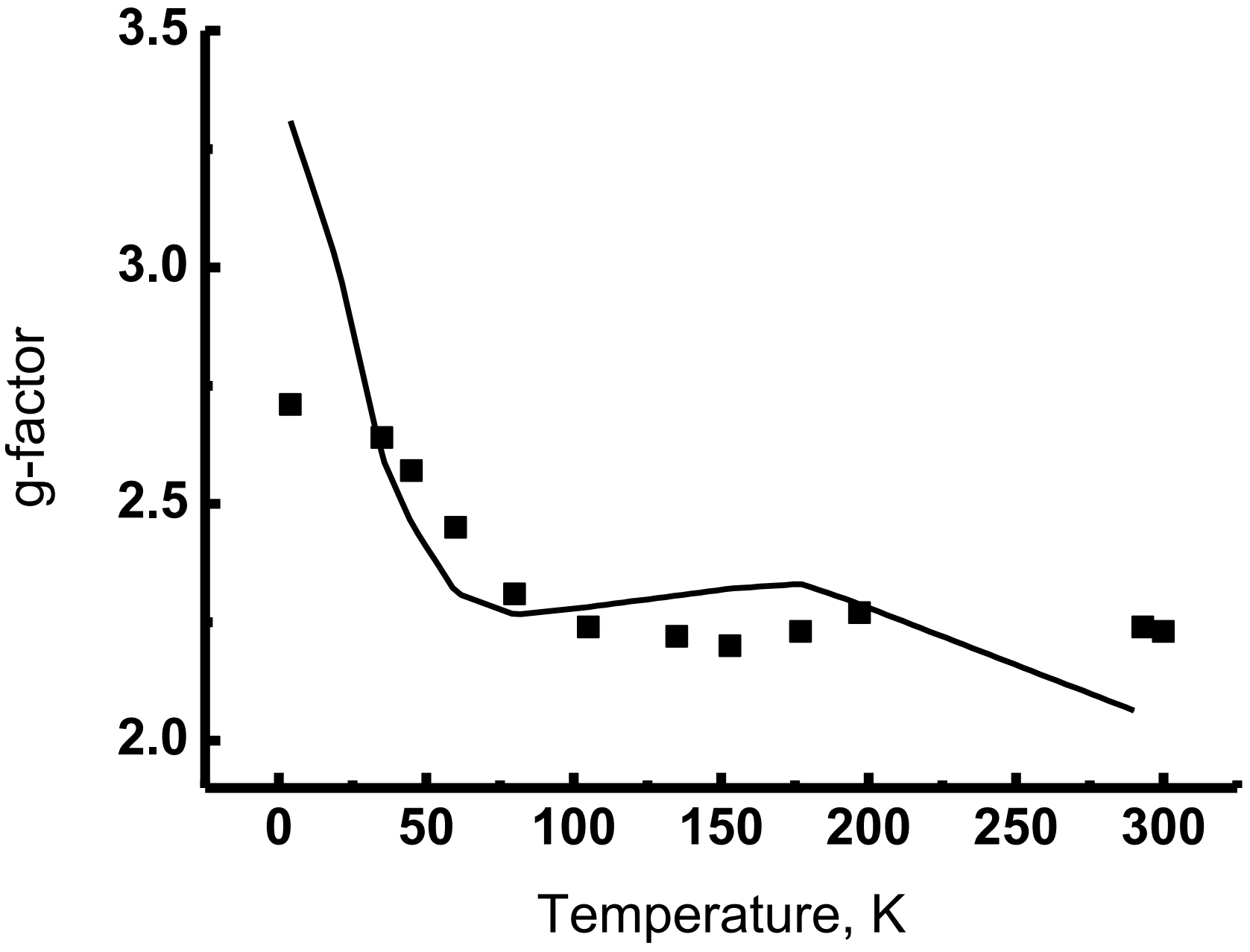

FIG. 4. $g$-factor dependence of the low-field EPR signal on temperature (points). The line is drawn according to Eq. (3).